\documentclass[pra,a4paper,twocolumn,showpacs,superscriptaddress]{revtex4}
\usepackage{amsmath}
\usepackage{amsfonts}
\usepackage{graphicx}
\usepackage{longtable}
\begin{document}

\title{Entropic uncertainty assisted by temporal memory} 
\author{H. S. Karthik}\email{karthik@rri.res.in}\affiliation{Raman Research Institute, Bangalore 560 080, India} 
\author{A. R. Usha Devi}\email{arutth@rediffmail.com} 
\affiliation{Department of Physics, Bangalore University, 
Bangalore-560 056, India}
\affiliation{Inspire Institute Inc., Alexandria, Virginia, 22303, USA.}
\author{J. Prabhu Tej}\email{j.prabhutej@gmail.com}
\affiliation{Department of Physics, Bangalore University, 
Bangalore-560 056, India}
\author{A. K. Rajagopal} \email{attipat.rajagopal@gmail.com}
\affiliation{Inspire Institute Inc., Alexandria, Virginia, 22303, USA.}
\affiliation{Harish-Chandra Research Institute, Chhatnag Road, Jhunsi, Allahabad 211 019, India.}
\date{\today}
\begin{abstract} 
The uncertainty principle brings out intrinsic quantum bounds on the precision of measuring non-commuting observables. Statistical outcomes in the measurement of incompatible observables reveal a trade-off on the sum of corresponding entropies. Massen-Uffink entropic uncertainty relation (Phys. Rev. Lett. {\bf 60}, 1103 (1988)) constrains the sum of entropies associated with incompatible measurements.  The entropic uncertainty principle in the presence of quantum memory (Nature Phys. {\bf 6}, 659 (2010)) brought about  a fascinating twist by showing that  quantum side information, enabled due to entanglement, helps in beating the uncertainty of non-commuting observables. Here we explore the interplay between temporal correlations and uncertainty.  We show that with the assistance of a prior quantum temporal information achieved by sequential observations on the same quantum system at different times, the uncertainty bound on entropies gets reduced.  
\end{abstract}
\maketitle

The uncertainty principle marks an astounding departure from  classical determinsm  by setting fundamental limits on the precision achievable in  {\em knowing}  non-commuting observables of a particle. To put it in simple terms, if one attempts to determine the position of a particle,  prediction of its momentum  gets  inaccurate. Heisenberg~\cite{Heisenberg} quantified this intrinsic quantum indeterminacy encrypted in the measurements of position and momentum in terms of  standard deviations $(\Delta Q)_\rho(\Delta P)_\rho\geq \frac{\hbar}{2}$ in any quantum state $\rho$ of the particle. 
Robertson~\cite{rob}  generalized it to arbitrary pairs of non-commuting observables $X$, $Z$  as 
\begin{equation} 
\label{rob}
(\Delta X)_\rho\, (\Delta Z)_\rho\geq \frac{1}{2}\,\vert\langle[X, Z]\rangle_\rho\vert.
\end{equation} 
Uncertainty relation (\ref{rob}) constraining the product of standard deviations suffers from the drawback that the right hand side of  (\ref{rob}) depends on the quantum state. In the specific example of a state  $\rho$ prepared in an eigenstate of $X$, 
the standard deviation  $(\Delta X)_\rho$ as well as the commutator  $\vert\langle[X, Z]\rangle_\rho\vert$ vanish and in turn, the uncertainty relation doesn't reveal any constraint on the {\em spread}  $(\Delta Z)_\rho$ of the observable $Z$. It has been identified subsequently that Shannon entropies of the probabilities of measurement outcomes of the observables $X$, $Z$ given by,     
$H_\rho(X)=-\sum_x \, P(x) \log_2 P(x)$, 
$H_\rho(Z)=-\sum_z \, P(z) \log_2 P(z)$  
offer a more general framework to quantify the {\em intrinsic  ignorance}  associated with incompatible measurements. Here,  $x\ (z)$ are the measurement outcomes  of the observable $X\ (Z)$ and $P(x)=\langle x\vert \rho\vert x\rangle$ ($P(z)=\langle z\vert \rho\vert z\rangle$) denote the probability of outcomes $x$ ($z$); $\{\vert x\rangle\} \  (\{\vert z \rangle\})$ is the set of eigenvectors 
of $X$ ($Z$). Trade-off between the entropies of a pair of discrete  non-commuting observables $X$ and $Z$ was formulated by Deutsch~\cite{Deutch} and was subsequently improved~\cite{partovi, bial2, kraus}. The conjecture put forth by Kraus~\cite{kraus} was proved by  Maassen and Uffink~\cite{MU}: 
\begin{equation} 
\label{mubound}
H_\rho(X)+H_\rho(Z)\geq - 2\log_2 c(X,Z),
\end{equation}
where $c(X,Z)={\rm max}_{x,y}\vert \langle x | z \rangle\vert$. The lower bound limiting the sum of entropies (\ref{mubound}) is independent of the state $\rho$. The term $c(X,Z)$ can assume a maximum value $\frac{1}{\sqrt{d}}$ resulting in the maximum entropic bound of $\log_2 d$,   where $d$ denotes the dimension of the system. When $\rho$ is an eigenstate of one of the observables, say $X$, the entropy of measurement $H_\rho(X)$ vanishes, but in turn the  entropy of the observable $Z$ gets constrained by $H_\rho(Z)\geq - 2\log_2 c(X,Z).$  

A recent uplifting happened  with the extension of entropic uncertainty relation assisted by the presence of a quantum memory~\cite{Berta}, which refined the lower bound of (\ref{mubound}). Here an observer Bob, whose task is to minimize the uncertainty of Alice's mesurement of the observables $X$, $Z$, is allowed to share an entangled quantum state $\rho_{AB}$  with that in Alice's possession. The uncertainty principle, when Bob possesses a quantum memory, is given by~\cite{Berta} 
\begin{equation} 
\label{bertabound}
S(X\vert B)+ S(Z\vert B)\geq -2\log_2 c(X,Z)+ S(A\vert B),
\end{equation}
where $S(X\vert B)=S(\rho^{(X)}_{AB})-S(\rho_B), S(Z\vert B)=S(\rho^{(Z)}_{AB})-S(\rho_B)$ are the conditional von Neumann entropies of the post measured states 
$\rho^{(X)}_{AB}=\sum_x(\Pi_x \otimes I_B) \rho_{AB} (\Pi_x \otimes I_B),  \ \ \rho^{(Z)}_{AB}=\sum_z(\Pi_z \otimes I_B) \rho_{AB} (\Pi_z \otimes I_B)$  obtained after the measurements of $X,\, Z$ performed by Alice on the system $A$;  $\Pi_x=\vert x\rangle\langle x\vert,\ \Pi_z=
\vert z\rangle\langle z\vert$; and  $S(A\vert B)=S(\rho_{AB})-S(\rho_B)$ is the conditional  von Neumann entropy of the state $\rho_{AB}$. When Alice's system is in a maximally entangled state with Bob's quantum memory, the second term on the right hand side of (\ref{bertabound}) takes negative value: $S(A\vert B)=-\log_2 d$ and as $- 2\log_2 c(X,Z)\leq \log_2 d$, one can acheive a trivial lower bound of zero.  Thus, with the help of a quantum memory maximally entangled with Alice's state, Bob can beat the uncertainty bound and can predict the outcomes of incompatible observables $X$, $Z$ precisely.  

Statistics of quantum correlations between the outcomes of spatially separated systems get  mimicked in an interesting fashion by that of temporally separated observables, measured sequentially in a single quantum system~\cite{lg,brukner,Vedral, uksr,emary}. Non-classicality of {\em temporal correlations} between outcomes of sequentially measured  observables is reflected by the violation of Leggett-Garg inequality~\cite{lg} (also termed as temporal Bell inequality~\cite{Mahler}), experimental verification of which has gained momentum recently~\cite{pala, mahesh, wald, souza, knee, katiyar}.  Sequential measurements  on the {\em same} quantum system result in the transmission of {\em temporal} information. Temporal correlations resulting from  consecutive observations on a single quantum system (in contrast to measurements on spatially separated systems) draw a  surge of interest  in  foundational investigations on quantum vs classical world view~\cite{Fritz, guhne}. Further, information gained from correlations between the outcomes of subsequent measurements on the same quantum system is shown to be advantageous in quantum communication tasks involving state discrimination~\cite{Hillery} and in quantum cryptography~\cite{Axta, Nori}.                   

In this letter, we raise the question, `analogous to spatial correlations, do temporal correlations arising in sequential measurement of observables, play a distinct role in reducing the uncertainty of incompatible observables?' This boils down to explore if the sum of conditional entropies $H_\rho(X\vert X_0)+H_\rho(Z\vert Z_0)$ is always smaller than the Massen-Uffink bound of $-2\log_2 c(X,Z)$ so, that  measurements of incompatible observables $X$, $Z$, conditioned by  outcomes of prior time measurements of $X_0$, $Z_0$ respectively,  lead to better precision.  We show that uncertainty does get reduced in the presence of a {\em quantum temporal memory} due to correlations between the outcomes of $X_0$ ($Z_0$) and $X$ ($Z)$ --  whereas it is impossible to beat the uncertainty bound  if the temporal correlations are {\em classical}.      

Let us consider a qubit prepared in a completely random mixture given by $\rho=I/2$ ($I$ denotes $2\times 2$ identity matrix). Measurements of the observables $X=\sigma_x$ and $Z=\sigma_z$ in this state leads to  Shannon entropies of measurement $H_\rho(X)=H_\rho(Z)=1$; $c(X,Z)=\frac{1}{\sqrt{2}}$ and the uncertainty bound (\ref{mubound}) is  $-2\log_2 c(X,Z)=1$; the Massen-Uffink relation is satisfied. Let us envisage the following scenario: A dichotomic observable $X_0=\cos\theta\ \sigma_z+\sin\theta\, \sigma_x$ is measured in the quantum state followed by which  $X=\sigma_x$ is sequentially measured; the probabilities of realizing the outcomes $x_0=\pm 1$ for $X_0$ and $x=\pm 1$ for $X$ in the sequential measurement is given by $P(x_0,x)={\rm Tr}[\Pi_{x_0}\rho\Pi_{x_0}\Pi_{x}]=\frac{1}{4}[1+x_0 x\, \cos\theta]$, where the projectors associated with dichotomic observables $X_0$ and $X$  are given by $\Pi_{x_0}=\frac{1}{2}[I+x_0 X_0]$, $\Pi_{x}=\frac{1}{2}[I+x X]$. Further, measurement of $Z=\sigma_z$ is preceded by that of another dichotomic observable $Z_0=\cos\phi\ \sigma_z+\sin\phi\ \sigma_x$ results in the probabilities  $P(z_0, z)={\rm Tr}[\Pi_{z_0}\rho\Pi_{z_0}\Pi_{z}]=\frac{1}{4}[1+z_0 z\, \cos\phi].$ The conditional Shannon entropy associated with the sequential measurement of $X_0$ and $X$  is given by  $H_\rho(X\vert X_0)=-\displaystyle\sum_{x_0,x=\pm1} P(x_0,x)\log_2 [P(x\vert x_0)]=\nobreak H[\cos^2(\theta/2)]$ (where the conditional probability $P(x\vert x_0)=P(x_0,x)/P(x_0$); $H(p)=-p\log_2 p-(1-p)\log_2(1-p);\ 0\leq p\leq 1$ denotes the binary entropy, which is bounded by $0\leq H(p)\leq 1$. Similarly, one gets the conditional Shannon entropy $H_\rho(Z\vert Z_0)=H[\cos^2(\phi/2)]$ associated with the sequential measurement of $Z_0$, $Z$. Clearly, the sum of conditional entropies  $H_\rho(X\vert X_0)+H_\rho(Z\vert Z_0)=H[\cos^2(\theta/2)]+H[\cos^2(\phi/2)]$ beats the uncertainty bound $-2\log_2 c(X,Z)=1$. More specifically, the uncertainty relation (\ref{mubound}) no longer holds for entropies of $X$ and $Z$ conditioned by the  information in the {\em temporal memory} obtained by prior measurements $X_0$, $Z_0$. While conditioning in general reduces the information entropy i.e., $H_\rho(X\vert X_0)\leq H_\rho(X), \ H_\rho(Z\vert Z_0)\leq H_\rho(Z)$,  we prove here that the temporal correlations between the sequential measurement outcomes of $X$, $X_0$  and $Z$, $Z_0$ must necessarily be  {\em non-classical} in order to beat the uncertainty bound of (\ref{mubound}), which operates in the absence of any temporal side information.

{\em Conditioned entropic uncertainty relation}: We proceed to prove the entropic uncertainty relation assisted by temporal correlations. Consider a single quantum system prepared in the state $\rho$. In the absense of any other assisting information, the uncertainty in the observables $X$ and $Z$ is bounded by (\ref{mubound}). A temporal memory is created by first noting down the outcome $x_0$ ($z_0$) of  an observable $X_0$ ($Z_0$) at an earlier time before recording the measurement outcomes $x$ ($z$)  of $X$ ($Z$). Then, the ignorance about the  measurement outcome of $X$ conditioned on  the information about $X_0$ stored in temporal memory is quantified in terms of the conditional Shannon entropy $H_\rho(X\vert X_0)$:   
\begin{eqnarray}
\label{hx0x}
H_\rho(X\vert X_0)&=& H_\rho(X_0, X)-H_\rho(X_0) \nonumber \\ 
&=& H_\rho(X)- H_\rho(X_0:X)
\end{eqnarray}                            
which is expressed in terms of the mutual information entropies $H_\rho(X_0:X)=H_\rho(X)+H_\rho(X_0)-H_\rho(X,X_0)$ and the unconditioned entropies $H_\rho(X)$. Similarly,  entropy of $Z$, conditioned by the outcomes of $Z_0$ is given by, 
\begin{equation}
\label{hy0y}
H_\rho(Z\vert Z_0)= H_\rho(Z)- H_\rho(Z:Z_0).
\end{equation}                            
The entropic uncertainty relation in the presence of temporal memory is then obtained by identifying the lower bound on the sum of conditional entropies 
$H_\rho(X\vert X_0)+H_\rho(Z\vert Z_0)$. Combining (\ref{hx0x}), (\ref{hy0y}), using the minimum value $\left[H_\rho(X)+H_\rho(Z)\right]_{\rm min}=-2\log_2 c(X,Z)$ (as given by (\ref{mubound})) and the maximum values $\left[H_\rho(X_0:X)\right]_{\rm max},\ \left[H_\rho(Z_0:Z)\right]_{\rm max}$ of the mutual information entropies,   we obtain, 
\begin{widetext}
\begin{eqnarray}
\label{tmEUR}
 H_\rho(X\vert X_0)+H_\rho(Z\vert Z_0)&\geq& {\rm max}\left[0, -2\log_2 c(X,Z)-\left\{H_\rho(X_0:X)\right\}_{\rm max}-\left\{H_\rho(Z_0:Z)\right\}_{\rm max}\right]\nonumber \\
&\geq & {\rm max}\left[0, -2\log_2 c(X,Z)-H^{({\rm \rm min})}_\rho(X)-H^{({\rm \rm min})}_\rho(Z)\right].
\end{eqnarray}          
\end{widetext}

The second line of the inequality (\ref{tmEUR}) follows by noting that  the mutual information entropy of two variables $X$, $X'$ can atmost be equal to the minimum of marginal entropies of $X$ or $X'$ i.e., 
$H(X:X')\leq{\rm min}[H(X), H(X')]$~\cite{NC}; denoting ${\rm min}[H_\rho(X), H_\rho(X_0)]=H^{({\rm \rm min})}_\rho(X)$, ${\rm min}[H_\rho(Z), H_\rho(Z_0)]=H^{({\rm \rm min})}_\rho(Z)$, we thus express $\left[H_\rho(X_0:X)\right]_{\rm max}=H^{({\rm \rm min})}_\rho(X)$ and  $\left[H_\rho(Z_0:Z)\right]_{\rm max}=H^{({\rm \rm min})}_\rho(Z)$. Further, since $-2\log_2 c(X,Z)$ cannot exceed $\log_2 d$ and the largest values of the marginal entropies $H^{({\rm \rm min})}_\rho(X), H^{({\rm \rm min})}_\rho(Z)$ being $\log_2 d$~\cite{NC}, the right hand side of the conditioned entropic uncertainty (\ref{tmEUR}) is expressed as the maximum of the trivial value zero and $-2\log_2 c(X,Z)-H^{({\rm \rm min})}_\rho(X)-H^{({\rm \rm min})}_\rho(Z)$.

{\em Conditioning with classical temporal correlations}: 
 Temporal correlation between the sequential outcomes $x_0$ and $x$ of the observables $X_0$, $X$ is {\em classical}~\cite{note} iff the joint probabilities $P(x_0, x)$ can be expressed as a convex combination of products of probabilities, 
\begin{eqnarray}
 \label{hvp}
 P(x_0, x)&=&\sum_\lambda\ p_\lambda\, P_\lambda(x_0)\, Q_\lambda(x), \\ 
 \label{hvm}
 \sum_{x_0} P_\lambda(x_0)&=& 1, \ \sum_{x} Q_\lambda(x)= 1 \\ 
 \sum_\lambda\, p_\lambda&=&1, \ \ 0\leq p_\lambda \leq 1.  \nonumber
\end{eqnarray} 
Quantum temporal memory requires that the correlation outcomes of the observables at different time instants are {\em not} governed by the joint probabilities of the form (\ref{hvp}).    

We now prove the following theorem.  

{\em Theorem}: If temporal correlations of the outcomes of $X_0$, $X$ and those of $Z_0$, $Z$ obtained from sequential measurement runs on the quantum state are {\em classical} (the correlation probabilities are of the form (\ref{hvp})), the sum of  conditional  entropies obey a {\em  temporal entropic steering inequality}~\cite{pra2013} 
\begin{equation}
H_\rho(X\vert X_0)+H_\rho(Z\vert Z_0)\geq -2\log_2 c(X,Z). 
\end{equation}      

{\em Proof}: Let us consider the conditional information for the measurement outcomes of the observable $X$, given that in a prior measurement  $X_0$ has taken the value $x_0$:  
\begin{equation} 
H_\rho(X\vert X_0=x_0)=-\sum_{x} P(x\vert x_0)\, \log_2 P(x\vert x_0)
\end{equation}      
The conditional probability $P(x\vert x_0)=P(x_0,x)/P(x_0)$ corresponding to {\em classical} temporal correlations (see (\ref{hvp})) is given by, 
\begin{eqnarray}
P(x\vert x_0)&=&\frac{\sum_\lambda p_\lambda\, P_\lambda(x_0)\, Q_\lambda(x)}{\sum_{\lambda'} p_{\lambda'}\, P_{\lambda'}(x_0)} \nonumber \\
 &=& \sum_\lambda p_{\lambda,x_0}\, Q_\lambda(x)
\end{eqnarray}
where we have denoted $p_{\lambda,x_0}=\frac{p_\lambda\, P_\lambda(x_0)}{\sum_{\lambda'} \, p_{\lambda'}\,P_{\lambda'}(x_0)}$. Note that $\sum_\lambda p_{\lambda,x_0}=1$, and 
$0\leq p_{\lambda,x_0}\leq 1$.

Consider the relative entropy $D({\cal P}_{x_0}\vert\vert {\cal Q}_{x_0})$ of the probabilitiy distributions ${\cal P}_{x_0}(\lambda, x)=p_{\lambda, x_0}\, Q_\lambda(x)$ and ${\cal Q}_{x_0}(\lambda, x)=p_{\lambda, x_0}\, P(x\vert x_0)$. Positivity of the relative entropy leads to the following identification~\cite{steering}: 
\begin{eqnarray}
D({\cal P}_{x_0}\vert\vert {\cal Q}_{x_0})&=&\sum_\lambda\, \sum_x\, p_{\lambda,x_0}\, Q_\lambda(x)\, \log_2\left[ \frac{Q_\lambda(x)}{P(x\vert x_0)}\right]\geq 0 \nonumber \\
&&\Rightarrow  H_\rho(X\vert x_0)\geq \sum_\lambda p_{\lambda, x_0}\ H^{(\lambda)}_\rho(X)
\end{eqnarray} 
where $H^{(\lambda)}_\rho(X)=-\sum_x Q_\lambda(x)\, \log_2 Q_\lambda(x)$.  
Thus, the average conditional information $H_\rho(X\vert X_0)=-\sum_{x_0}\ p(x_0)\ H_\rho(X\vert x_0);\ \  p(x_0)~=~\sum_x\, P(x, x_0)=\sum_\lambda p_\lambda\, P_\lambda(x_0)$ should obey the constraint 
\begin{eqnarray}
H_\rho(X\vert X_0)&\geq& \sum_{x_0} p(x_0)\,  \sum_\lambda p_{\lambda, x_0}\ H^{(\lambda)}_\rho(X) \nonumber \\   
&\geq& \sum_\lambda\ p_\lambda\  H_\rho^{(\lambda)}(X),  
\end{eqnarray}
Similarly, we obtain 
\begin{equation}
H_\rho(Z\vert Z_0)\geq  \sum_\lambda\ p_\lambda  H_\rho^{(\lambda)}(Z).  
\end{equation}
Thus, the sum of conditional entropies are constrained by 
\begin{eqnarray}
\label{mrEur}
H_\rho(X\vert X_0)+H_\rho(Z\vert Z_0)&\geq& \sum_\lambda\ p_\lambda  [H_\rho^{(\lambda)}(X)+H_\rho^{(\lambda)}(Z)]\nonumber \\
&=& -2\log_2\, c(X,Z).
\end{eqnarray}
In the second line of (\ref{mrEur}) we have employed the Massen-Uffink relation $H_\rho^{(\lambda)}(X)+H_\rho^{(\lambda)}(Z)\geq -2\log_2\, c(X,Z)$. 

This identification reveals the crucial significance of {\em quantum temporal memory} to achieve {\em sharpened} predictions of incompatible observables.         

\begin{figure}[h]
\includegraphics*[width=1.7in,keepaspectratio]{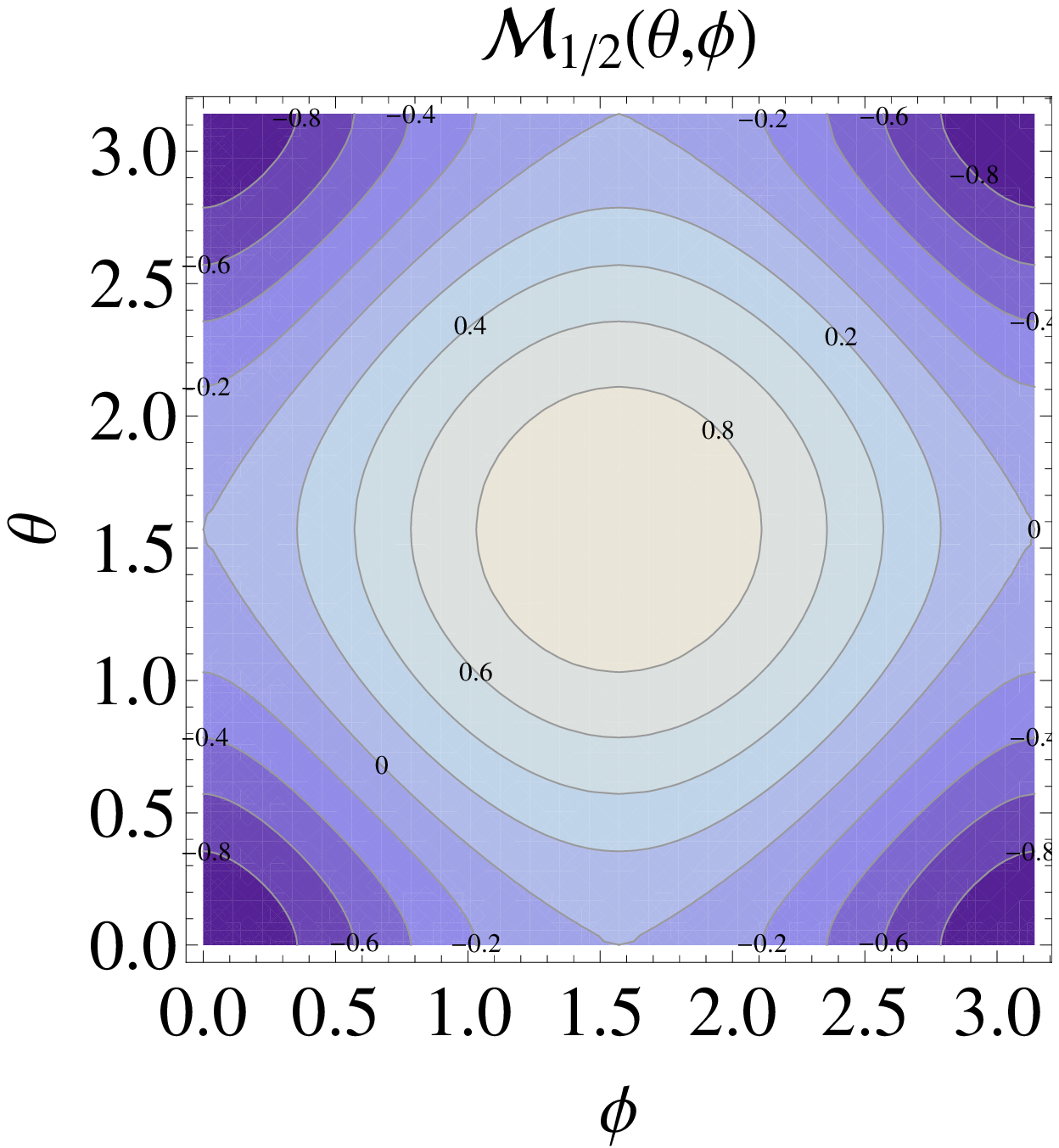}\includegraphics*[width=1.7in,keepaspectratio]{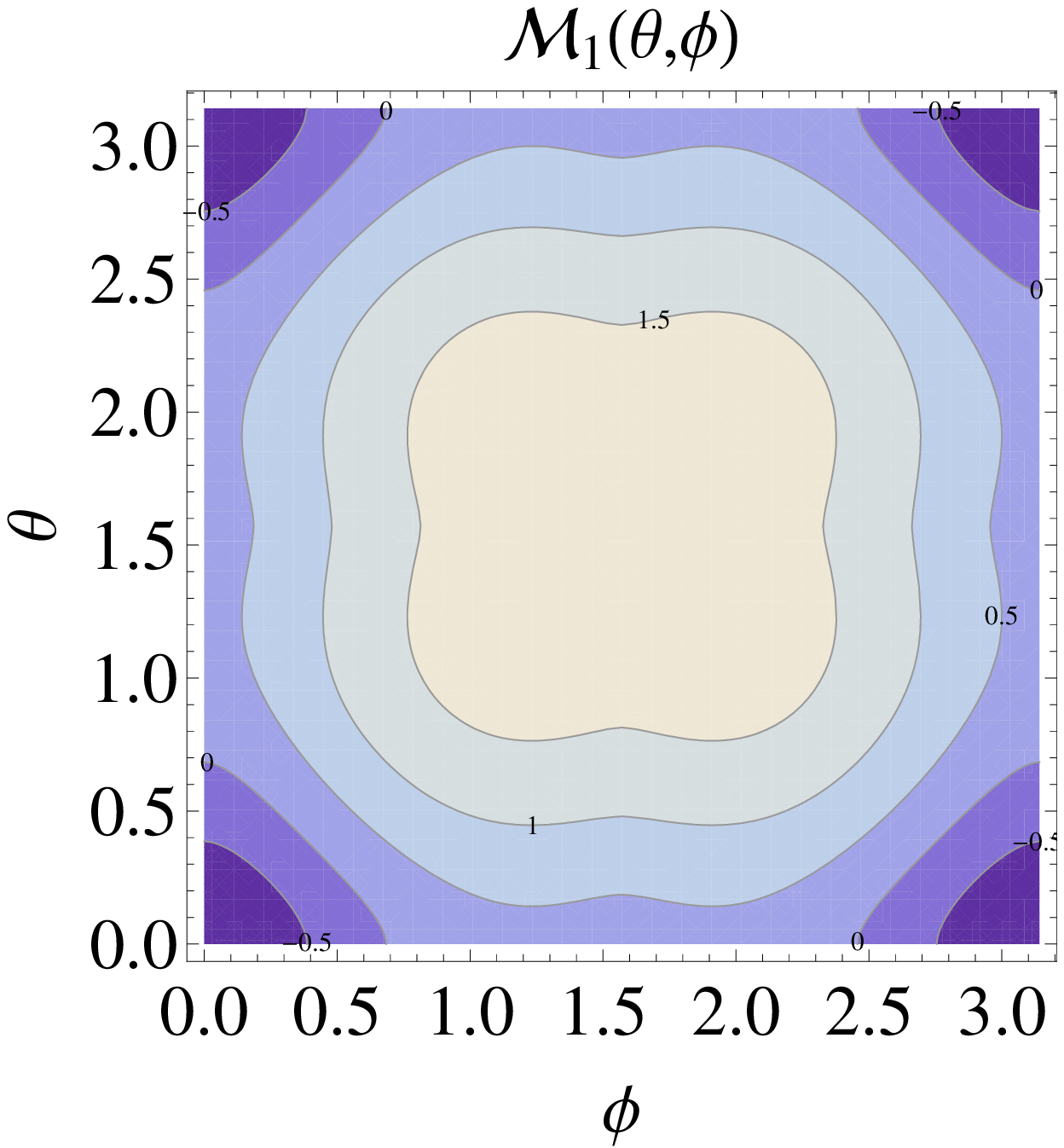}
\includegraphics*[width=1.7in,keepaspectratio]{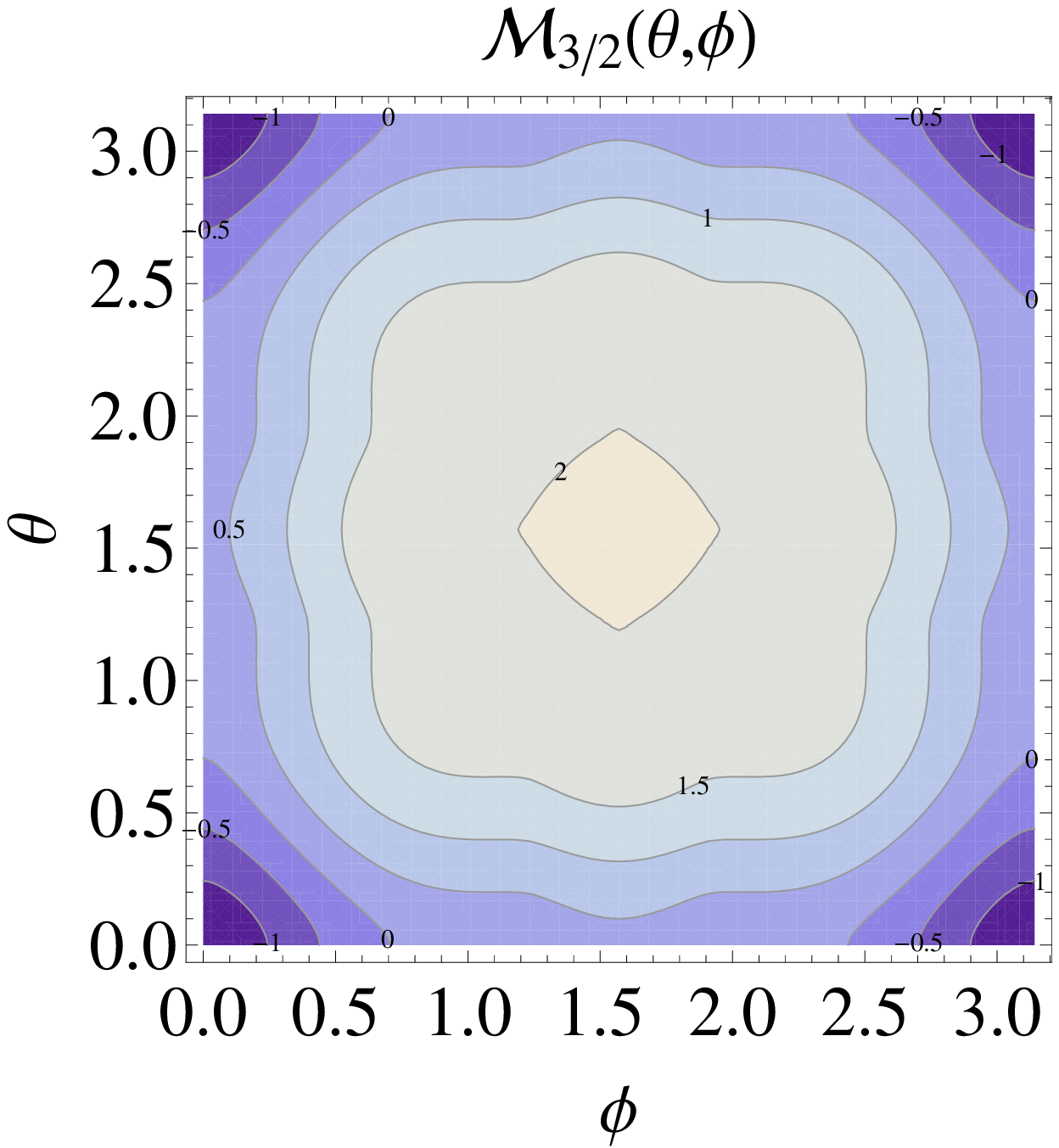}\includegraphics*[width=1.7in,keepaspectratio]{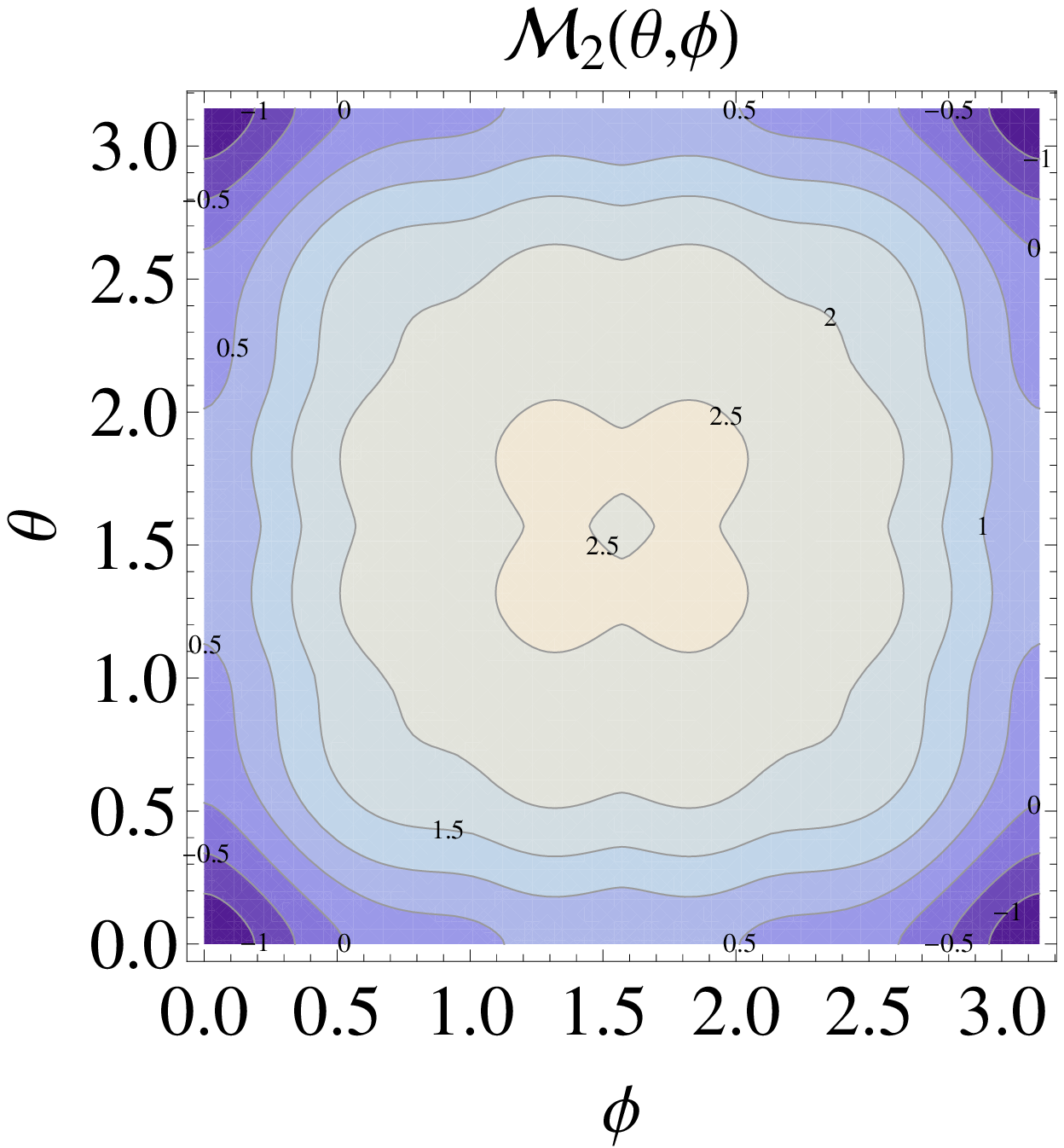}
\caption{(Color online) A contour plot of the quantity  ${\cal M}_s(\theta,\phi)$  (defined by (\ref{m})) constructed based on two runs of sequential measurements of  the spin component $S_z(t)$ of a quantum rotor, with dimensionless time separations $\theta$ and  $\phi$    for spin values $s=1/2$, $s=1$, $s=3/2$, $s=2$.   Negative values of ${\cal M}_s$ imply that the uncertainties about the outcomes of the spin components $S_x$, $S_z$ (conditioned on the information of outcomes of $S_z(t_{x0})$, $S_z(t_{z0})$ of prior measurements) get reduced in the presence of  quantum temporal memory. It may be seen that the range of values of the dimensionless time-separation parameters  $\theta$ and $\phi$, over which ${\cal M}_s$ is negative, reduces with the increase of spin $s$ indicating a quantum to classical transition of temporal correlations.  All quantities are dimensionless.}
\end{figure}

We illustrate how temporal correlations assist in reducing the entropic spread of non-commuting observables  by considering an example of a spin-$s$ quantum rotor prepared initially in a state $\rho=\frac{1}{2s+1}\sum_{m_z=-s}^{s}\ \vert s,m_z\rangle\langle s,m_z\vert =\frac{I_{2s+1}}{2s+1}$.      
(Here $\vert s, m_z\rangle$ are the simultaneous eigenstates of the squared spin operator $S^2=S_x^2+S_y^2+S_z^2$ and the $z$-component of spin $S_z$ (with respective eigenvalues $ s(s+1) $ and $m_z$); $I_{2s+1}$ is the $(2s+1)\times (2s+1)$ identity matrix). Measurement of non-commuting observables $X=S_x$ and $Z=S_z$ results in the probabilities of outcomes $-s\leq m_{x}, m_z\leq s$ as, $P(m_{x})={\rm Tr}[\rho\, \Pi_{m_x}]=\frac{1}{2s+1}; P(m_{z})={\rm Tr}[\rho\, \Pi_{m_z}]=\frac{1}{2s+1}$,  where $\Pi_m$ denotes the projection operator of the corresponding observable. The {\em spread} in the completely random measurement outcomes is revealed in terms of the corresponding Shannon entropies of measurement  $H_\rho(X)=\log_2 (2s+1)$ and $H_\rho(Z)=\log_2 (2s+1)$, which obey the trade-off relation (\ref{mubound}) --  the largest value of the uncertainty bound on the right hand side being $\log_2 (2s+1)$.  

In order to identify how entropic uncertainty relation for $S_x$ and $S_z$, assisted by prior conditioning, can reveal enhanced precision of the observables, we consider  dynamical evolution of the system  governed by the Hamiltonian ${\cal H}= \hbar\,\omega\, S_y$. Under the Hamiltonian dynamics, the evolution of $z$ component of spin is given by  $S_z(t)=e^{iS_y\omega t}\, S_z\, e^{-iS_y\omega t}=S_z\, \cos(\omega\, t)+S_x\, \sin(\omega\, t).$ We consider sequential measurement of $S_z(t)$ at different times as follows.  In the first run, the observable $S_z(t)$ is measured at time $t=t_{x0}$ and consequently at $t_x=\pi/2\omega$ (this corresponds to sequential measurement of observables $X_0=S_z\cos(\omega\, t_{x0})+S_x\sin(\omega t_{x0})$ and $X=S_x$) with a dimensionless time separation  $\theta=\omega t_{x0}-\pi/2$. The sequential measurements enable the observer to record the temporal correlation probabilities $P(m_{x_0}, m_x; \theta)$ of the outcomes $-s\leq m_{x_0}, m_x\leq s$ of the observables $X_0=S_z(t_0)$ and $X=S_x$. Then, in one more round of observations,    $S_z$ is measured sequentially at two different time instants $t_{z0}$ and $t_z=\pi/\omega$ (i.e., a measurement of $Z_0=S_z\cos(\omega\, t_{z0})+S_x\sin(\omega t_{z0})$ and then consequently $Z=S_z$), separated by a dimensionless time parameter $\phi=\omega\, t_{z0}-\pi$ is performed and the  correlation probabilities $P(m_{z_0}, m_z; \phi)$ of the $(2s+1)^2$ outcomes $-s\leq m_{z_0}, m_z\leq s$ are noted down.       

The probabilities of sequential measurement outcomes of $S_z(t)$ at two different times $t_{x0}$ and $t_{x}=\pi/2\omega$ are given by~\cite{uksr},  
\begin{eqnarray}
P(m_{x_0}, m_{x};\theta)=P(m_{x_0}; t_{x0})\, P(m_{x}; t_{x}\vert m_{x_0}; t_{x0}) \hskip 0.7 in\nonumber \\
 = {\rm Tr}[\rho\, \Pi_{m_{x_0}}(t_{x0})]\, \frac{{\rm Tr}[\Pi_{m_{x_0}}(t_{x0})\rho\, \Pi_{m_{x_0}}(t_{x0})\, \Pi_{m_{x}}(t_{x})]}{P(m_{x_0}; t_{x0})}\nonumber \\
= \frac{1}{2s+1}\, {\rm Tr}[\Pi_{m_{x_0}}(t_{x0})\, \Pi_{m_{x}}(t_{x})] \hskip 1.27in \nonumber \\ 
= \frac{1}{2s+1}\,  \vert \langle\, s, m_{x_0}\vert e^{-i\omega (t_{x0}-t_{x})\, S_y}\,\vert s, m_{x}\rangle\vert^2\hskip 0.70in\nonumber \\
=\frac{1}{2s+1}\,  \vert\, d^{s}_{m_{x}\, m_{x_0}}(\theta)\vert^2\hskip 1.8in
\end{eqnarray}  
where  $\Pi_{m}(t)=e^{i\omega t\, S_y}\, \vert s, m\rangle\langle s, m\vert\, e^{-i\omega t\, S_y}$ is the projection operator measuring the outcome $m$ of the spin component $S_z(t)$; and  $d^{s}_{m_x\, m_{x0}}(\theta)=\langle s, m_{x}\vert e^{-i\theta\, S_y}\vert s, m_{x_0}\rangle$ are the matrix elements of the $2s+1$ dimensional irreducible representation of rotation~\cite{Rose}  about $y$-axis by an angle $\theta=\omega (t_{x0}-t_{x})$. The marginal probability associated with measuring $S_z(t_{x_0})$ is readily obtained as   $P(m_{x_0}; t_{x_0})={\rm Tr}[\rho\, \Pi_{m_{x_0}}(t_{x0})]=\frac{1}{2s+1}$. Similarly, the correlation probabilties in the second run of sequential measurements are obtained as, 
$P(m_{z_0}, m_{z}; \phi)=\frac{1}{2s+1}\,  \vert\, d^{s}_{m_{z}\, m_{z_0}}(\phi)\vert^2$, 
and the marginal probabilities $P(m_{z_0}; t_{z_0})=1/(2s+1).$ The conditional entropies of measurement  (which depend only on the time separations $\theta, \ \phi$) $H_\rho(X\vert X_0)={\cal H}(\theta)$ and 
$H_\rho(Z\vert Z_0)={\cal H}(\phi)$   are given by,   
\begin{eqnarray}
{\cal H}(\theta)=-\frac{1}{2s+1}\, \sum_{m_{x_0}, m_{x}}\, \vert d^s_{m_{x_0},m_{x}}(\theta)\vert^2\, \log_2  \vert d^s_{m_{x_0},m_{x}}(\theta)\vert^2 \nonumber \\ 
{\cal H}(\phi)=-\frac{1}{2s+1}\, \sum_{m_{z_0}, m_{z}}\, \vert d^s_{m_{z_0},m_{z}}(\phi)\vert^2\, \log_2  \vert d^s_{m_{z_0},m_{z}}(\phi)\vert^2. \nonumber \\ 
\end{eqnarray}
We define a quantity ${\cal M}_s(\theta, \phi)$ as the difference between the sum of conditional entropies and the Massen-Uffink uncertainty bound $-2\log_2 c(X,Z)$ 
\begin{eqnarray}
\label{m}
{\cal M}_s(\theta, \phi)&=&H_\rho(X\vert X_0)+H_\rho(Z\vert Z_0)+2\, \log_2 c(X, Z)\nonumber \\
    &=& {\cal H}(\theta)+{\cal H}(\phi)+2\, \log_2 c(X, Z)
\end{eqnarray}                         
in order to demonstrate  improved precision in the measurement of the spin components $X=S_x$ and $Z=S_z$. While a {\em classical} temporal side information  results in ${\cal M}_s(\theta,\phi)$ being necessarily positive, presence of a {\em quantum temporal memory}, created by appropriate sequential measurements, can reveal itself in non-positive values of ${\cal M}_s(\theta,\phi)$. In Fig.~1, we have plotted the quantity ${\cal M}_s(\theta, \phi)$ for  as a function of  $\theta$ and $\phi$ for spin values $s=1/2, 1, 3/2$ and 2. The results clearly demonstrate reduction in the uncertainties of the non-commuting spin components $S_x$, $S_z$ (in the region where ${\cal M}_s$ is negative)  -- in the presence of a {\em quantum temporal memory}. We note that  the range of  time-separation  $\theta$ and $\phi$, over which  ${\cal M}_s$ assumes negative values, reduces with the increase of spin $s$ -- thus indicating a quantum to classical transition of  the temporal memory in the limit of large spin $s$.

{\em Conclusions:} Uncertainty principle reflects the inevitability inbuilt within the quantum framework in realizing deterministic outcomes for non-commuting observables of a particle.  Entropic uncertainty relation~\cite{MU} captures the trade-off in the {\em spread} of the outcomes of incompatible observables. However, a deterministic prediction is ensured  when the  particle is entangled maximally with another party. Berta et. al.,~\cite{Berta} brought out the subtle interplay between uncertainty and entanglement by extending the entropic uncertainty principle in the presence of quantum side information. In this work, we have explored the interesting association between temporal correlations and uncertainty. Our entropic uncertainty relation reveals that the presence of  quantum temporal side information too plays a significant role in beating the uncertainty bound.   More specifically, our results offer a unified view that a prior {\em quantum} knowledge, achieved with the help of suitable spatially/temporally separated observations, empower a determinstic prediction of non-commuting observables.


\begin{thebibliography}{0} 
\bibitem{Heisenberg} W. Heisenberg, W. 
Z. Phys. {\bf 43}, 172
(1927). 

\bibitem{rob} H. P. Robertson, 
Phys. Rev. {\bf 34}, 163-164 (1929).


\bibitem{bial1} I. Bialynicki-Birula and J. Mycielski, 
Commun. Math. Phys. {\bf 44}, 129
(1975).

\bibitem{Deutch} D. Deutsch, 
Phys. Rev. Lett. {\bf 50}, 631
(1983).


\bibitem{partovi}  M. H. Partovi,
Phys. Rev. Lett. {\bf 50}, 1883
(1983). 

\bibitem{bial2} I. Bialynicki-Birula, 
Phys. Lett. A {\bf 103} 253
(1984).

\bibitem{kraus} K. Kraus, 
Phys. Rev. D {\bf 35}, 3070
(1987).





\bibitem{MU} H. Maassen \& J. B. M. Uffink, 
Phys. Rev. Lett. {\bf 60}, 1103
(1988).

\bibitem{Berta} M. Berta, M. Christandl, R. Colbeck, J. M. Renes, and R. Renner, 
Nature Physics {\bf 6}, 659
(2010).

\bibitem{lg} A. J. Leggett and A. Garg, \prl {\bf 54}, 857 (1985).


\bibitem{brukner} C. Brukner, S. Taylor, S. Cheung, and V. Vedral, quant-
ph/0402127 (2004). 


\bibitem{Vedral} V. Vedral,  arXiv:1204.5559 (2012). 

\bibitem{uksr} A. R. Usha Devi, H. S. Karthik, Sudha, and A. K. Rajagopal,
Phys. Rev. A {\bf 87}, 052103 (2013).

\bibitem{emary} C. Emary, F. Nori, and N. Lambert, arXiv:1304.5133 (2013).

\bibitem{Mahler} J. P. Paz and G. Mahler, \prl {\bf 71}, 3235 (1993). 


\bibitem{pala} A. Palacios-Laloy, F. Mallet, F. Nguyen, P. Bertet, D. Vion, D. Esteve, and A. N. Korotkov, Nat. Phys. {\bf 6}, 442 (2010). 

\bibitem{mahesh} V. Athalye, S. S. Roy, and T. S. Mahesh, \prl {\bf 107}, 130402 (2011). 

\bibitem{wald} G. Waldherr, P. Neumann, S. F. Huelga, F. Jelezko, and J. Wrachtrup, \prl {\bf 107}, 090401 (2011). 

\bibitem{souza} A. M. Souza, I. S. Oliveira, and R. S. Sarthour, New. J. Phys. {\bf 13}, 053023 (2011). 

\bibitem{knee} G. C. Knee et al., Nat. Comm. {\bf 3}, 606 (2012). 

\bibitem{katiyar} H. Katiyar, A. Shukla, K. R. K. Rao, and T. S. Mahesh, Phys. Rev. A {\bf 87}, 052102 (2013).
 
\bibitem{Fritz} T. Fritz, New J. Phys. {\bf 12}, 083055 (2010).

\bibitem{guhne} C. Budroni, T. Moroder, M. Kleinmann, and O. G\"uhne, \prl {\bf 111}, 020403 (2013).

\bibitem{Hillery} J. Bergou, E. Feldman, and M. Hillery, \prl {\bf 111}, 100501 (2013).

\bibitem{Axta} H. Akshata Shenoy, S. Aravinda, R. Srikanth, and D. Home,  arXiv:1310.0438 (2013). 

\bibitem{Nori} Y. Chen, C. Li, N. Lambert, S. Chen, Y. Ota, G. Chen, and F. Nori,  arXiv:1310.4970 (2013). 

\bibitem{NC} M. A. Nielsen and I. L. Chaung, {\em Quantum Computation and
Quantum Information} (Cambridge University Press, Cambridge,
2002).    

\bibitem{note} {\em Classicality} of temporal correlations rests on the assumption that the observables have definite pre-existing values and measurement of an observable at a given instant has no consequence on its subsequent evolution~\cite{lg}. 

\bibitem{pra2013} In the context of spatially separated parties sharing a quantum state $\rho_{AB}$, the inequality $H_{\rho_{AB}}(X_A\vert X_{0B})+H_{\rho_{AB}}(Z_A\vert Z_{0B})\geq -2\log_2 c(X,Z)$ (where Alice measures $X$ ($Z$) conditioned by the outcome Bob obtains for $X_0$ ($Z_0$) on the state in his possession) is  referred to as entropic Einstein-Podolsky-Rosen steering inequality.  See   
J. Schneeloch, C. J. Broadbent, S. P. Walborn, E. G. Cavalcanti, and J. C. Howell, \pra {\bf 87}, 062103 (2013). 

\bibitem{steering} This identification is similar to the one outlined by S. P. Walborn, A. Salles, R. M. Gomes, F. Toscano, and P. H.
Souto Ribeiro, \prl {\bf 106}, 130402 (2011).

\bibitem{Rose} M. E. Rose, {\em Elementary theory of angular momentum}, (John Wiley, New York, 1957) 


\end{thebibliography}
\end{document}